\documentstyle[11pt,fleqn,cite]{article}   
\catcode`\@=11
\@addtoreset{equation}{section}
\def\theequation{\arabic{section}.\arabic{equation}}
\def\appendix{\renewcommand{\thesection}{\Alph{section}}\setcounter{section}{0}
              \renewcommand{\theequation}
            {\mbox{\Alph{section}.\arabic{equation}}}\setcounter{equation}{0}}
\def\maketitle{\thispagestyle{empty}\setcounter{page}0\newpage
                \renewcommand{\thefootnote}{\arabic{footnote}}
                  \setcounter{footnote}0}
\renewcommand{\thanks}[1]{\renewcommand{\thefootnote}{\fnsymbol{footnote}}
               \footnote{#1}\renewcommand{\thefootnote}{\arabic{footnote}}}

\renewcommand{\title}[1]{\begin{center}\Large\bf #1\end{center}\rm\par\bigskip}
\renewcommand{\author}[1]{\begin{center}\Large #1\end{center}}
\newcommand{\address}[1]{\begin{center}\large #1\end{center}}

\def\dinfn{\smallskip Dipartimento di Fisica, Universit\`a di Trento\\ 
                           and Istituto Nazionale di Fisica Nucleare,\\
                                   Gruppo Collegato di Trento, Italia}

\def\Idinfn{\address{\dinfn}}

\newcommand{\email}[1]{e-mail: \sl #1@science.unitn.it\rm}

\newcommand{\femail}[1]{\thanks{\email{#1}}}
\newcommand{\pacs}[1]{\smallskip\noindent{\sl PACS numbers:
                       \hspace{0.3cm}#1}\par\bigskip\rm}
\def\babs{\hrule\par\begin{description}\item{Abstract: }\it} 
\def\eabs{\par\end{description}\hrule\par\medskip\rm}
\renewcommand{\date}[1]{\par\bigskip\par\sl\hfill #1\par\medskip\par\rm}
\newcommand{\ack}[1]{\par\section*{Acknowledgments} #1} 
\newcommand{\s}[1]{\section{#1}}
\renewcommand{\ss}[1]{\subsection{#1}}

\newcommand{\ca}[1]{{\cal #1}}         
\def\hs{\qquad}               
\def\nn{\nonumber}            
\def\beq{\begin{eqnarray}}    
\def\eeq{\end{eqnarray}}      
\def\R{{\hbox{{\rm I}\kern-.2em\hbox{\rm R}}}}   
\def\H{{\hbox{{\rm I}\kern-.2em\hbox{\rm H}}}}   
\def\N{{\hbox{{\rm I}\kern-.2em\hbox{\rm N}}}}   
\def\C{{\ \hbox{{\rm I}\kern-.6em\hbox{\bf C}}}} 
\def\Z{{\hbox{{\rm Z}\kern-.4em\hbox{\rm Z}}}}   
\def\ii{\infty}                                  
\renewcommand{\Re}{\mathop{\rm Re}\nolimits}       
\renewcommand{\Im}{\mathop{\rm Im}\nolimits}       

\def\be{\beta}

\def\de{\delta}
\def\ep{\varepsilon}

\def\si{\sigma}

\def\th{\theta}

\def\Ga{\Gamma}

\def\La{\Lambda}
\def\Si{\Sigma}

\def\Th{\Theta}

\begin{document}

\title{Black holes with unusual topology}

\author{L.~Vanzo \femail{vanzo}}

\Idinfn

\date{may 1997}

\begin{abstract}
The Einstein's equations with a negative cosmological constant admit 
solutions which are asymptotically anti-de Sitter space. Matter fields 
in anti-de Sitter space can be in stable equilibrium even if the potential 
energy is unbounded from below, violating the weak energy condition. 
Hence there is no fundamental reason that black hole's horizons should 
have spherical topology. In anti-de Sitter space the 
Einstein's equations admit black hole solutions where the horizon can be 
a Riemann surface with genus $g$. The case $g=0$ is the asymptotically 
anti-de Sitter black hole first studied by Hawking-Page, which has 
spherical topology. The genus one black hole has a new free parameter 
entering the metric, the conformal class to which the torus belongs. 
The genus $g>1$ black hole 
has no other free parameters apart from the mass and the charge. All 
such black holes 
exhibits a natural temperature which is identified as the period of 
the Euclidean continuation and there is a mass formula connecting the 
mass with the surface gravity and the horizon area of the black hole. 
The Euclidean action and entropy are computed and used to argue that 
the mass spectrum of states is positive definite.  
\end{abstract} 

\pacs{04.20.-q, 04.70.Bw, 04.70.Dy}

\maketitle

\s{Introduction}

In general relativity it was widely believed that black holes formed by 
gravitational collapse should have spherical horizon \cite{hawk73b}. In the 
stationary case this is ensured by Hawking's theorem 
\cite{hawk72-25-152}, under the assumption of asymptotic flatness and 
positivity of matter energy. The "topological censorship theorem" of 
Friedmann, Schleich and Witt \cite{frie93-71-1486} is another 
indication of the impossibility of non spherical horizons. The theorem 
states that in a globally hyperbolic, asymptotically flat spacetime, 
any two causal curves extending from past to future null 
infinity are homotopic. As pointed out by Jacobson and 
Venkataramani \cite{jaco95-12-1055}, a black hole with toroidal surface 
topology would provides a possible violation of topological censorship, 
as a light ray from past 
infinity linking with the hole of the torus and then back to future 
infinity would not be deformable to a light ray traveling from past 
to future outside the black hole. Thus the hole must quickly close up, 
before a light ray can pass through. In fact, as was shown by Shapiro, 
Teutolsky and Winicour \cite{shap95-52-6982}, a temporarily toroidal 
horizon can form in gravitational collapse, in a way consistent with 
the theorems. For non stationary black holes, and under the 
assumptions of asymptotic flatness and the dominant energy condition 
for matter fields, Gannon \cite{gann76-7-219} proved that a smooth 
black hole must be either a two-sphere or a torus. All these results 
made essential use of the condition of asymptotic flatness, which 
entails a vanishing cosmological constant.

The Einstein's equations with cosmological term, $\La$, admit black 
hole solutions which are asymptotic to either de Sitter ($\La>0$) or 
anti-de Sitter ($\La<0$) space. 
These solutions have spherical horizon and obey thermodynamics laws like 
asymptotically flat black holes 
\cite{hawk77-15-2738,hawk83-87-577,brow94-50-6394}.  
In de Sitter space, one can find locally static solutions of the form
\beq
ds^2=-Vdt^2+V^{-1}dr^2+r^2d\si^2\hs V=C-\frac{2m}{r}-\frac{\La 
r^2}{3}\nn
\eeq 
for any $C$, provided the two dimensional line element $d\si^2$ 
has constant curvature $k=2C$. Then for $C>0$ we have the 
asymptotically de Sitter black hole, with positive mass and spherical 
horizons. If $C<0$, the black hole interpretation of the solution is 
lost unless the mass parameter is negative. In anti-de Sitter space the 
situation is just the opposite. In $2+1$-dimensions there are 
the recently discovered locally anti-de Sitter black hole solutions 
\cite{teit92-69-1849}, which have constant 
curvature everywhere not just asymptotically at infinity, and the 
Brill's multi-black hole's solution \cite{bril96-69-249}. The horizon 
of a $2+1$-dimensional spacetime is a closed line, which leaves not much 
space for introducing non trivial topology. On the other hand, there do not 
seem to exist a reasonable, higher dimensional generalization of 
the BTZ's black hole. The metrics recently found have horizons with spherical 
topology, but all the anti-de Sitter conserved charges are infinite
\cite{bana97-grqc}. So apparently, one had to give up the condition 
of constant curvature. Planar and cylindrical black holes in anti-de 
Sitter space were indeed discovered by Lemos 
\cite{lemo95-12-1081,lemo96-353-46}, which upon compactification 
became toroidal. Open and closed black strings \cite{stro91-360-197} 
are also likely to form topologically toroidal black holes 
\cite{lemo96-54-3840}. 
On the other hand, \AA minneborg et al. \cite{peld96-13-2707} presented a 
class of solutions in $3+1$-dimensions, displaying the causal 
structure characteristic of black holes, and having constant negative 
curvature everywhere. Hence they was locally isometric to anti-de Sitter 
space but, surprisingly, showed an event horizon with the topology of a 
Riemann surface with arbitrary genus. 
Finally, R.~Mann \cite{mann97-14-L109}, and then D.~Brill in 
collaboration with J.~Louko \cite{bril96-69-249}, introduced 
a class of black solutions admitting all the above horizon 
topologies, which can have both positive or negative mass, 
which can be charged, and which have 
a curvature singularity in the origin\footnote{Recently, the author also 
met the uncharged version of Mann' solution \cite{vanzo97-grqc}.}. At the moment 
it was unclear whether these topological black holes could result from 
gravitational collapse but, since then, this question was also settled 
affirmatively \cite{mann97-grqc}. So we finally have topologically non 
trivial black holes, albeit in anti-de Sitter space. 

Although anti-de Sitter space does not seem to 
correspond to the world in which we live, its importance has been 
noticed in many occasions 
\cite{fron65-37-221,avis78-18-3565,cole80-21-3305,brei82-144-249,filt91-43-485,abbo82-195-76}.
Two features seem worth mentioning. Firstly, anti-de Sitter and Weyl 
conformal gravity are the only type of gravity which have a consistent 
interaction with massless higher spin fields \cite{frad87-177-63} and, 
secondly, consistent anti-de Sitter strings exist for any $D\neq26$ 
(or $D\neq10$) \cite{frad91-261-26}, provided the cosmological term 
has the critical value which is required by anomaly cancellation. 

In this paper we would like to investigate the thermodynamics 
properties of the topological black holes from the point of view of 
the Euclidean formulation (for a detailed treatment of the Hamiltonian 
thermodynamics of asymptotically anti-de Sitter black holes see 
\cite{louk96-54-2647}). We point out that higher genus black holes are 
really "cosmological black holes", inasmuch as their size is the size of 
the (anti-de Sitter) universe itself. Hence they could only exist during the 
inflationary era, when the cosmological constant was not small. The 
toroidal black hole, on the other hand, can exist in a virtually flat 
space, as the size is governed by the mass and the conformal class of 
the torus, rather than by the cosmological constant. 
 
In Sec.~(I), we begin by presenting the metric and discussing the relevant 
geometric features, including the asymptotic behaviour at infinity. We 
shall not discuss entirely the causal structure (it is presented in 
\cite{bril97-grqc}), nor we make it confident 
how the black hole could result from gravitational 
collapse of some, topologically non trivial (i.e. non spherical) 
matter configuration (this is explained in \cite{mann97-grqc}). In 
Sec.~(II), we define the mass and show it obeys a 
Smarr-like formula. We point out that, due to the asymptotic behaviour of the 
metric, there is no way to make finite the Hamiltonian than 
subtracting a reference background in the same topology class of the 
actual solution. The natural choice would seem to be the solutions of 
\AA minneborg et al., to which the black hole approaches asymptotically, but a 
thermodynamics argument will favor a rather different choice. 
In Sec.~(III), we determine the off-shell Euclidean action and use it to 
evaluate the entropy of the black hole. Some 
discrepancies regarding the mass spectrum will then be resolved. 

In the following, we shall use the curvature conventions of 
Hawking-Ellis's book \cite{hawk73b} and employ Planck dimensionless units.  

\ss{The topological black holes}

The class of metrics to consider is
\beq
ds^2=-Vdt^2+V^{-1}dr^2+r^2\si_{ij}dx^idx^j
\eeq
where $\si_{ij}$ is the metric of a two-manifold, $S$, which is not 
assumed to be a topological sphere, and $V=f(r)$. The non 
vanishing components of the Ricci tensor are 
\beq
R_{tt}=-V^2R_{rr}=\frac{1}{2}V\,V^{''}+\frac{VV^{'}}{r}\hs
R_{ij}=\ca R_{ij}-(rV^{'}+V)\si_{ij}
\eeq
where the calligraphic's $\ca R_{ij}$ refers to $\si_{ij}$. Now one 
verifies immediately that the function
\beq
V=\kappa-\frac{k^{'}}{r}+\frac{r^2}{\ell^2}
\eeq
makes the metric to satisfy Einstein's equations with negative 
cosmological constant, $R_{ab}=\La g_{ab}$, $\La=-3\ell^{-2}$, 
for any pair $(\kappa,k^{'})$. The surprising fact is that for 
this to be true, the two dimensional metric $\si_{ij}$ must satisfy 
the equations for a constant curvature surface, which need not  
be a sphere, namely $\ca R_{ij}=\kappa\si_{ij}$ and $\ca R=2\kappa$. 
Therefore if $\kappa=-q^2<0$, the two-manifold $S$ must be a 
surface with constant, negative curvature. If this surface is compact and 
orientable, then it must be a Riemann surface of genus $g>1$ for 
$q^2>0$. If $q=0$, then the surface is a torus, and $q=\pm i/R$ gives 
a sphere of radius $R$. Actually, the parameter $q$ is 
fictitious as long as non zero, since we can always rescale $t$, 
$r$, $k^{'}$ and $\si_{ij}$ so as to achieve that $q=1$. Hence we take 
the metric of the uncharged, genus-$g$, black hole in the form  
\beq
ds^2=-\left(-1-\frac{2\eta}{r}+\frac{r^2}{\ell^2}\right)dt^2+
\left(-1-\frac{2\eta}{r}+\frac{r^2}{\ell^2}\right)^{-1}dr^2+
r^2\si_{ij}dx^idx^j
\label{mebh}
\eeq
where now $\ca R_{ij}=-\si_{ij}$ describes a Riemann surface with 
genus $g>1$ and Euler number $\chi_g=2-2g$. In the genus one case, we 
pick a complex number $\tau$, with $\Im\tau>0$ (this is known as the 
Teichm\"{u}ller complex parameter of the torus). Such a complex number 
specify a class of conformally equivalent tori, two tori being 
equivalent if and only if the respectives Teichm\"{u}ller parameters 
are connected by a fractional linear transformation with integer 
coefficients. We shall write the flat metric of the torus in the form
\beq
d\si^2=\si_{ij}dx^idx^j=|\tau|^2dx^2+dy^2+2\Re\tau dx\,dy
\eeq
where the pair $(x,y)$ ranges over the closed unit square in $\R^2$. 
The toroidal, uncharged black hole metric is now
\beq
ds^2=-\left(-\frac{2\eta}{r}+\frac{r^2}{\ell^2}\right)dt^2+
\left(-\frac{2\eta}{r}+\frac{r^2}{\ell^2}\right)^{-1}dr^2+
r^2(|\tau|^2dx^2+dy^2+2\Re\tau dx\,dy)
\label{mebhT}
\eeq
Let $\de(a,b)=1$ for $a=b$ and zero otherwise. 
From the Gauss-Bonnet theorem, the area of $S$ is
\beq
\ca A=-2\pi\chi_g+|\Im\tau|\de(g,1)=4\pi(g-1)+|\Im\tau|\de(g,1)
\eeq 
The metric possesses an irremovable 
singularity at $r=0$, because the invariant $R_{abcd}R^{abcd}$ blows 
up like $r^{-6}$ near $r=0$. Therefore, in the following, we shall study 
the metric for $r>0$ only.

We consider now whether the space represents a genuine black 
hole. The standard procedure to analyze black holes is to investigate the 
causal structure. In the $g>1$ case, the lapse function of the metric 
(\ref{mebh}) always has a real root at some $r_+$. This is the 
solution of the cubic 
equation $r^3-\ell^2\,r-2\eta\ell^2=0$, and the character of the roots 
depends on the sign of the discriminant, $\ca D=\eta^2\ell^4-\ell^6/27$. 
If $\ca D>0$ and $\eta>0$ then
\beq
r_+=\frac{2^{1/3}\ell^2}{3[2\eta\ell^2+2(\ca D)^{1/2}]^{1/3}}+
\frac{[2\eta\ell^2+2(\ca D)^{1/2}]^{1/3}}{2^{1/3}}
\label{root1}
\eeq
is the only real root, the singularity is spacelike and 
hidden inside an event horizon. If $\ca D>0$ but $\eta<0$ there is 
one negative real root, the lapse function is positive in the range 
$r>0$, and $r=0$ is a naked singularity. 
If $\ca D<0$, the allowed range for $\eta$ is 
$-\ell/3\sqrt{3}\leq\eta\leq\ell/3\sqrt{3}$. If $\eta>0$, there is one 
positive root which can be written as
\beq
r_+=\frac{2\ell}{\sqrt{3}}\cos(\th/3)\hs 
\cos\th=\frac{3\sqrt{3}\eta}{\ell}
\label{root2}
\eeq
where $\th\in[0,\pi/2]$, the other two roots being real and negatives. 
Again the singularity is spacelike and hidden. 
If $\eta<0$ there are two positive roots, $r_+$ and $r_-$, with 
$r_+>r_-$, corresponding to the 
choices $\th/3$ and $(\th+4\pi)/3$ in Eq.~(\ref{root2}), given 
$\th\in[\pi/2,\pi]$, and one negative root. 
Again the greater root represents an event 
horizon, and the region in between the two positive roots resembles 
the Reissner-Nordstr\"{o}m solution. In the region $0<r<r_-$ the lapse 
function is positive, so the singularity is timelike and $r=r_-$ 
represents an inner Cauchy horizon. The structure of this black hole is 
then quite complex. As we shall see, $\eta$ is related to the mass of the 
black hole, hence 
what we have here is a putative, negative mass black hole with an acceptable 
causal structure, the allowed range of "negative mass" being 
$M>-\ell/3\sqrt{3}$. At last, $\eta=-\ell/3\sqrt{3}$ gives a naked 
singularity and the solution has no black hole interpretation. It 
corresponds to the extreme limit where the inner horizon has the same 
location as the outer horizon, $r_-=r_+$, and it will play an 
important role when developing the thermodynamics Euclidean theory. 
Finally, there is a case whereby $\ca D=0$, or $\ell^2=27\eta^2$, for which 
again there is only one positive root at $r_+=6\eta$, the other two being 
equal but negatives. The genus one case is simpler, as the only positive root 
is at $r_+=(2\eta\ell^2)^{1/3}$. 

In all cases, the root $r_+$ makes the hypersurfaces 
$r=r_+$ an event horizon. The metric admits a Kruskal like extension in which the 
$r=0$ singularity is spacelike (as in the Schwarzschild solution), the 
reason being that the lapse function changes sign by crossing the 
horizon, except when $\eta<0$, in which case the singularity is timelike. 
Because of this fact, each future directed null geodesic behind the 
horizon will inevitably crash into the singularity at $r=0$, so it can 
never reach infinity. A related fact is that the expansion of each 
$r=$constant surface, with $r<r_+$, is negative and as such it 
is a closed trapped surface. The solution therefore represents a black hole 
for all $\eta>-\ell/3\sqrt{3}$, for $g>1$, or for all positive $\eta$ 
if $g=1$. The horizon has a portion to the future of the static region 
$r>r_+$, and a portion to the past. The two sheets intersect in a 
genus-$g$ Riemann surface, which is the fixed point set of the time 
translation symmetry of the solution. The horizon is thus a bifurcate 
Killing horizon. It has a surface gravity, $\kappa_g$ for the 
genus-$g$ case, which can be computed by standard means as
\beq
\kappa_g=\frac{3r_+^2-\ell^2}{2r_+\ell^2}\hs \kappa_1=\frac{3r_+}{2\ell^2}
\label{surfg}
\eeq
The surface gravity is non negative and 
vanishes only for the extreme solution, when $\eta=-\ell/3\sqrt{3}$ and 
$r_+=\ell/\sqrt{3}$. The area section of the horizon is 
\beq
A=4\pi r_+^2(g-1)+\de(g,1)\,r_+^2|\Im\tau|
\eeq
where $\tau$ is the Teichm\"{u}ller parameter of the torus. The mass 
of the black hole is, unlike the geometry, a rather delicate matter, 
and we shall discuss this question after having analyzed few 
asymptotics property of the metric.

To understand the geometrical origin of the genus-$g$ surfaces 
, let us pause for a moment with the black hole and 
consider the solution with $\eta=0$ and $\kappa=-q^2$ not, normalized to 
$-1$. The curvature tensor for this solution is 
$R_{abcd}=-\ell^{-2}[g_{ac}g_{bd}-g_{ad}g_{bc}]$, 
which shows that the space is locally isometric to the universal 
covering of anti-de Sitter space. The surprise comes when computing 
the curvature tensor of the $r=$constant surfaces. 
It is given by $\ca R_{ijkl}=-q^2[\si_{ik}\si_{jl}-\si_{il}\si_{jk}]$, 
and therefore it describes a space 
of constant, negative curvature again. Anti-de Sitter space, 
AdS for short, is the maximally symmetric space which is obtained by
restricting the metric $ds^2=-dx^2-dv^2+dy^2+dz^2+du^2$ in $\R^5$, 
with rectangular coordinates $(x,v,y,z,u)$, to the hyperboloid 
\beq
-x^2+y^2+z^2+u^2-v^2=-\ell^2
\eeq
The cosmological constant figuring in Einstein's equations 
is $\La=-3\ell^{-2}$. The topology of the space is 
that of $S^1\times\R^3$, but notice that 
each circle $x^2+v^2=\tau^2$ gives a closed timelike curve in AdS. Hence 
we pass to the covering by opening the circle into a real line. Given 
this, we note that by fixing $v^2-u^2=\ell^2\xi^2$ to be greater 
than $\ell^2$, i.e. $\xi^2>1$, makes the 
three remaining coordinates to range over hyperbolic two-space, 
which we denote by $H^2$. The orbits of constant $\xi$ describe uniformly 
accelerated observers in anti-de Sitter space, and we shall see 
now that the remaining $H^2$, which carries a positive definite 
metric, is the acceleration horizon of such observers. To this aim, we 
make use of the following parametrization of the hyperboloid
\begin{eqnarray}
x&=&\ell\sqrt{1+q^{-2}\xi^2}\cosh\rho \\
y&=&\ell\sqrt{1+q^{-2}\xi^2}\sinh\rho\cos\th \\
z&=&\ell\sqrt{1+q^{-2}\xi^2}\sinh\rho\sin\th \\
u&=&q^{-1}\ell\xi\cosh(qt/\ell) \\
v&=&q^{-1}\ell\xi\sinh(qt/\ell)
\end{eqnarray}
and then set $r^2=\ell^2(q^2+\xi^2)$. The induced metric takes the form
\beq
ds^2=-\left(-q^2+\frac{r^2}{\ell^2}\right)dt^2+\left(-q^2+\frac{r^2}{\ell^2}
\right)^{-1}dr^2+r^2d\si^2
\label{adsm}
\eeq
where $d\si^2=q^{-2}[d\rho^2+\sinh^2\rho d\th^2]$ is one of the many forms in 
which the metric of hyperbolic two-space $H^2$ is presented. Setting 
as before $q^2=1$, the metric differs from Eq.~(\ref{mebh}) by the 
absence of the crucial term $2\eta/r$, but is otherwise identical. 

The lapse function of the metric has a zero at $r_+=\ell$, which 
makes the metric of the three surface $r=r_+$ degenerate. This surface 
is in fact a bifurcate  event horizon, the future portion intersecting 
the past portion in a transverse $H^2$, which is the fixed point set 
ot the time translation symmetry. Although the metric displays the 
properties of a black hole, it is not in fact, as it represents the 
portion of AdS which is causally accessible to a family of accelerated 
observers. This is not the end of the story, as $H^2$ 
is non compact and we want a compact horizon. The $SO(2,3)$ symmetry 
group of AdS contains an $SO(1,2)$ subgroup acting on the $(x,y,z)$ 
sector of the five coordinates. This symmetry leaves unaffected the 
accelerated trajectories and only mixes the points in $H^2$, where it 
acts as a group of isometries. It is a well known fact that any 
Riemann surface with genus $g>1$ is the quotient space of $H^2$ by a 
discrete subgroup of isometries (roughly speaking, this is a subgroup 
whose elements can be labeled by an integer), acting in $H^2$ 
without fixed points (including infinity in $H^2$, so for example 
discrete translations are forbidden) and properly discontinuously 
(this means that the translates of any compact set are disjoints). 
Thus we may pick up such a discrete subgroup, say $\Ga$, and make the 
quotient (i.e. the orbit space). This makes the horizon a compact 
Riemann surface of genus $g>1$. The genus one case 
apparently has not such interpretation, nevertheless it can also be 
obtained identifying points in AdS space \cite{peld96-13-2707} and 
the metric is Eq.~(\ref{adsm}) with $q^2=0$. As we shall see, 
thermodynamics 
arguments indicate that this solution has positive mass, even in the 
absence of the $2\eta/r$ term in the metric. We shall call the resulting 
spacetime the RadS (Riemann-anti-de Sitter space), and we conclude that 
this is the asymptotic region of the topological black holes.

\ss{The mass and size of the black holes}

As is well known, there is a certain amount of freedom in defining 
the mass of the black hole, as this involves the subtraction of a zero 
point of energy. Looking at the metric (\ref{adsm}), it would seem 
natural to define the mass by taking RadS as a reference background, 
which has $\eta=0$, even if its topology is not that of anti-de Sitter 
space. However, for $g>1$ and for reasons to be explained below, we prefer to 
take as a reference background a metric in the class given by 
Eq.~(\ref{mebh}), with a "mass parameter" $\eta_0$. We shall also denote 
all quantities referring to the background with a subscript "$0$". The 
two values of $\eta_0$ we will discuss are then $\eta_0=0$ and 
$\eta_0=-\ell/3\sqrt{3}$, which is the lowest possible value for the 
metric to admit a black hole interpretation, $\eta\leq\eta_0$ being a 
naked singularity. In 
the case $g=1$, the background will be the metric (\ref{mebhT}), but with 
$\eta_0=0$, which again is the lowest value for the 
metric to admit a black hole interpretation, $\eta_0<0$ being a naked 
singularity.
 
We shall now identify the mass of the black hole as the on-shell value 
of the Hamiltonian, with lapse function $N=\sqrt{V}$ and vanishing shift 
vector 
\cite{regg74-88-286,york93-47-1407,hawk96-13-1487,louk96-54-2647}. 
To this aim, one puts a 
timelike boundary at same large $r=R$ and uses the Hamiltonian of 
general relativity in a manifold with boundary, taking care of all the 
boundary terms. At the end, one takes 
the limit as $R$ goes to infinity. As the $t$=constant slices are 
orthogonal to the timelike boundary at large distances 
which contains the Killing observers at "infinity", 
there are no "corner" terms in the Hamiltonian \cite{hayw93-47-3275}, 
and the mass is
\beq
M=-\frac{1}{8\pi}\int_{S_g}\sqrt{V}(\Th-\Th_0)r^2\sqrt{\si}\,d^2x 
\label{hamm}
\eeq
where $S_g$ is a Riemann surface with genus $g$, $\Th$ is the trace of 
the extrinsic curvature of $S_g$ as embedded in a $t=$constant 
hypersurface, $\Th_0$ is the same quantity as if $S_g$ were 
embedded in the reference spacetime, and the limit $R\rightarrow\ii$ is 
understood. The 
trace $\Th$ can be computed as the covariant divergence of the normal 
vector field to the boundary at $r=R$,  $\xi^a=\sqrt{V}\de^a_1$, in 
the induced metric of the $t$=constant slices , $h_{ab}=g_{ab}+u_au_b$, 
where $u^a$ is the future pointing normal. It is  
\beq
\Th=\frac{2}{R}\sqrt{-1-\frac{2\eta}{R}+\frac{R^2}{\ell^2}}
\simeq_{R\rightarrow\ii}\,\frac{2}{R}\sqrt{-1+\frac{R^2}{\ell^2}}+
\frac{2\eta}{R^2}\frac{1}{\sqrt{-1+\frac{R^2}{\ell^2}}}
\eeq
Similarly
\beq
\Th_0\simeq_{R\rightarrow\ii}\,\frac{2}{R}\sqrt{-1+\frac{R^2}{\ell^2}}
+\frac{2\eta_0}{R^2}\frac{1}{\sqrt{-1+\frac{R^2}{\ell^2}}}
\eeq
Therefore, asymptotically, $N(\Th-\Th_0)\simeq 2(\eta-\eta_0)R^{-2}$. One can 
repeat the calculation with the torus black hole metric, finding 
again the same result. From Eq.~(\ref{hamm}), in a condensed notation 
for any genus $g$, we obtain
\beq
M=-(\eta-\eta_0)\chi_g/2+\frac{\eta A}{4\pi r_+^2}\,\de(g,1)=
(\eta-\eta_0)(g-1)+\frac{\eta|\Im\tau|}{4\pi}\,\de(g,1)
\eeq
We see that even if the "$-1$" in the lapse functions $N$, $N_0$ does 
not count asymptotically, the integration over the boundary must 
involve a Riemann surface in the background with the same genus of the 
actual solution. The topology of the background must then be 
asymptotically $\R\times S_g$, with $S_g$ carrying a constant negative 
curvature metric (as required by Einstein's equations). This cannot be 
embedded in flat space (because then the curvature could not be 
negative everywhere), nor in the anti-de Sitter slices, which have 
topology $\R\times S^2$. If the background is to be a static solution of 
Einstein's equations, then presumably the metric (\ref{adsm}) is the 
only one available which has no curvature singularities, and the metric 
(\ref{mebh}) with $\eta_0=-\ell/3\sqrt{3}$ is the only one which has 
no black hole interpretation, although it has a naked singularity. 
The metric (\ref{mebh}) makes sense even for $\eta<0$ 
\cite{mann97-grqc,mann97-05grqc}, but again it has a curvature 
singularity in the origin. The $\eta_0=0$ background would seem 
preferable, as demanded by 
the topology of spacetime and by the absence of curvature singularities. 
We shall argue differently, however, when discussing the Euclidean theory.  

In the following, it will be convenient to parametrize the reference 
background by $r_0$ rather than $\eta_0$, $r_0$ being the positive root of the 
cubic equation $r^3-\ell^2r-2\eta_0\ell^2=0$. To the value $\eta_0=0$ 
corresponds then $r_0=\ell$ and to the critical value 
$\eta_0=-\ell/3\sqrt{3}$ corresponds the critical value 
$r_0\equiv r_c=\ell/\sqrt{3}$. This a 
double root of the lapse function, at which the background surface 
gravity vanishes. Finally, the toroidal background has $\eta_0=0$ and 
$r_0=0$.

If around the black hole there is a matter distribution with 
stress-energy tensor $T_{ab}$, then one can obtain a mass formula 
along the lines of \cite{hawk73-31-161}, by integrating the Killing 
identity
\beq
\nabla^a\nabla_bK_a=R_{bc}K^c 
\eeq
for the Killing field $K^a$, over a spacelike hypersurface $\Si$, 
which is asymptotically orthogonal 
to the trajectories of the Killing observers at infinity and intersects the 
horizon in a two-surface $S_g$. To this we must subtract, in addition, the 
volume contribution of the background with the same genus (both the 
solution and the background have a divergent vacuum energy 
, because $\La\neq0$). This has an horizon at $r=r_0$ and a surface 
gravity, $\kappa_0$. The mass formula reads
\beq
M=\frac{\kappa 
A}{4\pi}-\frac{\kappa_0A_0}{4\pi}+\frac{A}{4\pi\ell^2r_+^2}
(r_0^3-r_+^3)+\int_{\Si}(2T_{ab}-Tg_{ab})K^au^b\,d^3x 
\label{massf}
\eeq
the third term being the difference of the volume vacuum energy in the 
solution and the background. The mass so defined is also equal to the 
Abbott-Deser mass for asymptotically AdS spaces \cite{abbo82-195-76}, 
if only one repeats their analysis in the present case, and satisfy 
the first law for any $g$, which in the vacuum reads
\beq
dM=\frac{\kappa dA}{8\pi}
\eeq
The knowledge of the mass allows one to 
obtain some result about the size of the topological black holes. 
The radius of the black hole as seen from the outside static region is 
the value, $r_+$, of the real positive root of the lapse. This depends 
on a single parameter, $\eta$, that we showed is related to the black 
hole mass. The scale of the solution is determined by the 
cosmological constant, or by $\ell$ (present estimates would put a value 
for $\ell$ not less than $10^{27}\div10^{28}$ cm, which is about the 
size of the observable universe). For $g>1$, according to Eq.~(\ref{root1}), 
the black hole can have any size for masses grater than $\ell/\sqrt{27}$ 
and grows like $r_+\sim M^{1/3}$ for $M\ell\gg1$. This seems to be 
the less interesting case for large values of $\ell$. The degenerate 
case $\ca D=0$, is a black hole with $r_+=6\eta\sim M\sim\ell$, and 
the size of the black hole is the size of the universe. There is finally 
a case whereby $\ca D<0$. The mass of the black hole is bounded by a 
number of order $\ell$, the greater the mass the greater is the size, 
but this is always of order $\ell$. Hence there are no "small" 
topological, $g>1$ black holes, except for large values of the 
cosmological constant. In fact, the minimum size is $2\ell/\sqrt{3}$ 
for $\ca D\geq0$ and $\ell/\sqrt{3}$ if $\ca D<0$. If the black hole can radiate 
away its mass without changing the topology, then this would leave 
behind a cosmological horizon with finite size or a naked singularity. 
The toroidal black hole is more promising, as then 
$r_+=(2\eta\ell^2)^{1/3}$. As a function of the mass this is
\beq
r_+=\left(\frac{8\pi\ell^2 M}{|\Im\tau|}\right)^{1/3}
\eeq
and thus it depends on the conformal class of the torus. Now small 
black holes can exist with any mass and, within a given conformal 
class, they can exist for arbitrary large values of $\ell$. If the 
cosmological constant, though exceedingly small, is nevertheless 
finite, the toroidal black hole could exist in a virtually flat space. 

\ss{Euclidean formulation}

One approach to the thermodynamics of 
black holes, is to analyze the Euclidean action which one obtains 
under Wick rotation to imaginary time \cite{hawk77-15-2738}. 
The Euclidean black hole solution is obtained by rotating the 
time coordinate to imaginary values, and exist as a Riemannian 
metric for $r>r_+$. In the Euclidean section of the metric, the imaginary 
time plays the role of an angular coordinate, where the rotation 
"axis" is just the horizon. Therefore the metric will have 
a conic singularity at $r=r_+$, unless the imaginary time is identified 
with the right periodicity, which is
\beq
\be_+=\frac{2\pi}{\kappa}=[1-\de(g,1)]
\,\frac{4\pi\ell^2r_+}{3r_+^2-\ell^2}+
\de(g,1)\,\frac{4\pi\ell^2}{3r_+}
\eeq
An important exception to this is the critical solution with 
$\eta_0=-\ell/3\sqrt{3}$. This is the only solution for which 
the imaginary time can be 
identified with any period without loosing the regularity of the 
metric. This fact will have important consequences for the Euclidean 
theory. Unlike the asymptotically AdS black hole studied in 
\cite{hawk83-87-577,brow94-50-6394}, the period has no maximum value and is 
never zero, so the solution exist for any $\be_+$. However, 
$\be_+>2\pi\ell$ corresponds to negative energy states, if the 
prescription $\eta_0=0$ is adopted. The temperature of 
the genus $g$ black holes is therefore, for $g>1$ and $g=1$ 
respectively
\beq
T=\be_+^{-1}=\frac{3r_+^2-\ell^2}{4\pi\ell^2r_+}\hs
T=\frac{3r_+}{4\pi\ell^2}
\label{temp}
\eeq
The quantum origin of this temperature is hidden here by our choice of 
units. The identification of the period in imaginary time (a 
classical concept) with the inverse temperature of the equilibrium 
state, has no classical analog since the required Wick rotation is 
really $t\rightarrow-i\hbar\be$. That $T$ is a temperature can also be 
seen from the fact that one can construct the analog of the 
Hartle-Hawking quantum state as well as the analog of the Unruh state 
(work in preparation). To define the former, one imposes the boundary 
condition that the ingoing and outgoing fluxes of radiation, from and 
to timelike infinity, be equal. Both states have a temperature which is 
zero at infinity due to infinite redshift. However, zero rest mass 
particles escape to infinity arranged in a thermal flux with the 
temperature $T$, but their "angular distribution" is governed by the 
eigenfuctions of the Laplace operator on a Riemann surface rather than 
by the spherical harmonics.

The mass of the black hole as a function of the temperature is an 
important thermodynamics input. It can be obtained from the mass 
formula by expressing $r_+$ in terms of $T$ using Eq.~(\ref{temp}), 
which gives, for $g>1$ and $g=1$ respectively
\beq
r_+=\frac{2\pi\ell^2T}{3}\left[1+\sqrt{1+
\frac{3}{4\pi^2\ell^2T^2}}\right]\hs r_+=\frac{4\pi\ell^2T}{3}
\eeq
For $g>1$, the mass is a rather complicated function of this temperature 
\begin{eqnarray}
M\!&=&\!\frac{(g-1)4\pi^3\ell^4T^3}{27}\left(1+\sqrt{1+
\frac{3}{4\pi^2\ell^2T^2}}\right)\left[\left(2+\frac{3}{4\pi^2\ell^2T^2}+
2\sqrt{1+\frac{3}{4\pi^2\ell^2T^2}}\right)\!-\ell^2\right]\nn \\
&-&\eta_0(g-1)\geq-\left(\frac{\ell}{3\sqrt{3}}+\eta_0\right)(g-1)
\label{mass}
\end{eqnarray}
This mass is an increasing function of $T$, with a large-$T$ behaviour 
$M\sim T^3$, in the full range $0\leq T\leq\ii$ and the zero 
temperature state is a black hole with mass 
$M=-(\eta_0+\ell/3\sqrt{3})(g-1)$. The mass then increases till the 
temperature reaches the value $T=1/2\pi\ell$, at which the mass 
is $M= -\eta_0(g-1)$. The first prescription, $\eta_0=0$, gives then a massless 
black hole at finite temperature, at the end of a continuous spectrum 
of negative energy states, and the second prescription gives a continuous 
positive mass spectrum, although at this stage the terminology is 
conventional. However, it would seem natural to call "ground state" 
the state with zero temperature. For $g=1$ the mass is
\beq
M=|\Im\tau|\frac{8\pi^2\ell^4}{27}\,T^3
\label{mass1}
\eeq
Hence the stability condition, $\partial M/\partial T>0$, 
is fulfilled in every case. We shall now 
compute the off-shell Euclidean action of the black hole  
\beq
I=-\frac{1}{16\pi}\int_{\ca M}(R-2\La)\sqrt{g}\,d^4x-\frac{1}{8\pi}
\int_{\partial\ca M}K\sqrt{h}\,d^3x
\label{actoff}
\eeq
where $\partial\ca M=S^1\times S_g$ is the boundary of the solution 
identified with period $\be\neq\be_+$ at 
some fixed $r=R$, which will be taken to infinity at the end, and 
$K$ is the trace of the extrinsic curvature of the boundary. 
The Euclidean action so defined is a divergent function of 
the boundary location, and therefore it will be necessary to 
subtract from it the Euclidean action of a chosen background. For 
black holes which are asymptotically flat, de Sitter or anti-de 
Sitter, one can compute the difference of the Euclidean action of the 
actual solution with that of flat space, the four-sphere or the 
four-dimensional hyperbolic space respectively, these spaces being the 
Euclidean sections of the lorentzian metrics. In flat space, the 
Euclidean action comes entirely from the difference in the surface 
terms, the four-sphere has no boundary and the action 
is already finite without subtractions 
\cite{hawk77-15-2738,gibb77-15-2752}, in hyperbolic four-space the 
surface integral of the solution cancel the 
surface integral of the background and the action comes again from the 
difference in four volumes \cite{hawk83-87-577}. 

In the present case, we have apparently no other choice than comparing 
the Euclidean action of the black hole with that of another solution 
in the same topological class (i.e same Euler number). This is because 
with no other topology will the metric cancel the divergences coming from 
the surface and volume terms in the action, without fine tuning the 
parameters. For example, choosing anti-de Sitter requires fine tuning 
of the cosmological constant of the background to achieve 
cancellation of the leading divergences. Another choice could be taking a 
background in the same topology class but with a metric chosen by hand 
to cancel divergences. In general, however, this will not be a solution of 
Euclidean Einstein's equations, the procedure appears a little bit 
arbitrary and, moreover, the mass was defined relative to a specific 
background. 

Therefore we shall compute the difference between the 
Euclidean action of the black hole and that of a background in the same 
topology class, and for off-shell values of the inverse 
black hole's temperature. To agree with the mass definition, the 
$\eta_0$-parameter of the background will be either zero or $-\ell/3\sqrt{3}$.
In doing so, one encounters a conical 
singularity in the solutions as well as the background, except in 
the second case. We notice 
that such singularity in the background would persist even for on-shell 
values of $\be$, as the natural period of the background is different 
from $\be_+$. To compute the effect of the conical singularity one 
cuts, out of the manifold, a small disk around the horizon (in the 
Euclidean black hole this is an axis of rotation) at $r=\ep$, and then 
compute separately the action in the volume from $r=\ep$ to $r=R$, 
and the disk. The contribution of the disk is given, 
as is well known\cite{teit94-51-4315,suss94-50-2700,furs96-54-2711}, 
by the Gauss-Bonnet theorem and is
\beq
\frac{1}{16\pi}\int R\sqrt{g}d^4x=\frac{A}{4\be_+}(\be_+-\be)
\eeq
where $A$ is the area of the event horizon. The background contributes 
the same quantity or zero, depending on whether $r_0=\ell$
, for the choice $\eta_0=0$, or $r_0=r_c=\ell/\sqrt{3}$ for the choice 
$\eta_0=-\ell/3\sqrt{3}$. In the former case, $A_0=4\pi\ell^2(g-1)$ 
takes the place of $A$ and $\be_0=2\pi\ell$ the place of $\be_+$. 
Finally, the conic contribution of the toroidal background vanishes, 
too. Since for both metrics $R=4\La$, the volume's difference in the 
action of the two metrics is 
\beq
\frac{\be \ca A}{8\pi\ell^2}(R^3-r_+^3+r_0^3-R_0^3)\nn
\eeq
wher $\ca A=4\pi(g-1)+\de(g,1)|\Im\tau|$ and $R_0$ is the radial 
coordinate of the boundary in the background 
metric. This must be matched to $R$ by requiring the two metrics to agree 
asymptotically, which gives (for the torus is $\eta_0=0$)
\beq
R_0=R-\frac{(\eta-\eta_0)\ell^2}{3R^2}+O(R^{-3})\nn
\eeq  
up to terms of higher order in $R^{-1}$. Finally there is the surface 
contribution, which involves the asymptotic of the extrinsic 
curvatures in the form (this is the integrand of the boundary term in 
Eq.~(\ref{actoff}), after subtraction)
\beq
R^2NK-R_0^2N_0K_0=R^2\left(\frac{2V}{R}+\frac{V^{'}}{2}\right)-
R_0^2\left(\frac{2V_0}{R_0}+\frac{V_0^{'}}{2}\right)
\eeq
where $N=\sqrt{V}$ and $N_0=\sqrt{V_0}$ are the lapse functions of the 
black hole and the background respectively. Using the matching 
condition one finds this to vanish at infinity, and therefore the 
surface term also vanishes. Finally, we eliminate $\eta$ in favor of the 
mass and we obtain the following off-shell Euclidean action, valid for 
any genus $g$ (we recall the Kronecker symbol $\de(a,b)=1$ if $a=b$ and 
zero otherwise)
\beq
I=\frac{A}{4\be_+}(\be-\be_+)+[1-\de(r_0,r_c)]
\frac{A_0}{4\be_0}(\be_0-\be)+
\frac{\be A(r_0^3-r_+^3)}{8\pi\ell^2r_+^2}+\frac{\be M}{2}
\eeq
Notice that the conic contribution of the background is absent if 
$r_0=r_c=\ell/\sqrt{3}$, which correspond to the zero temperature 
state, or what is considered a "negative mass" solution in 
\cite{mann97-grqc,bril97-grqc}. Using the mass formula 
Eq.~(\ref{massf}), one can write the action in the form 
\beq
I=\be M-\frac{A}{4}+[1-\de(r_0,r_c)]\frac{A_0}{4}
\label{acnew}
\eeq
From this formula it would seem that a mass shift, though moving 
the negative energy states to positive values, would leave unaffected 
the entropy. This is wrong, because shifting $M$ by 
$\eta_0=-\ell/3\sqrt{3}$, removes at the same time the conic 
singularity in the background, and consequently affects the entropy. 
The on-shell action is $I$ evaluated at $\be=\be_+$, that is
\beq
I=[1-\de(r_0,r_c)]\frac{A_0}{4\be_0}(\be_0-\be_+)+
\frac{\be_+A(r_0^3-r_+^3)}{8\pi\ell^2r_+^2}
+\frac{\be_+M}{2}
\eeq
As a function of the black hole's temperature and for $g>1$, this is
\beq
I&=&\frac{M}{2T}-\frac{4\pi^3\ell^4T^2(g-1)}{27}\left[1+
\sqrt{1+\frac{3}{4\pi^2\ell^2T^2}}\right]^3
+\frac{(g-1)r_0^2}{2\ell}\left(\frac{r_0}{\ell}-
\frac{2\pi\ell}{\be_0}\right)T^{-1}\nn \\
&+&[1-\de(r_0,r_c)]\frac{A_0}{4}
\label{acT}
\eeq
where $M$ is given by Eq.~(\ref{mass}), and for the torus is
\beq
I=-|\Im\tau|\frac{4\pi^2\ell^4T^2}{27}
\label{acT1}
\eeq
The $T$ behaviour is exactly $-T^2$ for the torus and $-T^2$ 
asymptotically for higher genus, as for a massless boson gas in two 
spatial dimensions. In the tree approximation one identifies 
$I=-\log Z(\be)$, the 
partition function of the black hole \cite{gibb77-15-2752}. The 
density of states is the inverse Laplace transform of the partition 
function
\beq
\rho(E)=\frac{1}{2\pi i}\int_{\Re\be=c}Z(\be)e^{\be E}d\be
\eeq
For large energy the integral is dominated by the small-$\be$ limit of 
the partition function, which is of order $\exp(-C\be^{-2})$, $C>0$. 
Then the integrand has a saddle point at 
\beq
\be\simeq(2C/E)^{1/3}
\eeq
at which the second derivative of the logarithm of the integrand is 
positive. The path of steepest descent is then parallel to the 
imaginary axis and the integral gives
\beq
\rho(E)\simeq\exp(C_1E^{2/3})\hs C_1>0
\eeq
for some computable new constant $C_1$. For genus one the result is exact, 
but in every case the exponent is precisely $A/4$. This growth of 
$\rho(E)$ with $E$ makes it evident the existence of the partition 
function from the point of view of the "sum over states". The 
stability of the canonical ensemble is proved in \cite{bril97-grqc}, 
using the method of the reduced action, and is indicated by the 
positivity of the specific heat.  

Adopting $\log Z(\be)=-I$ as the partition function, where $I$ is the 
off-shell Euclidean action, allows one to evaluate the expectation value 
of the energy in the canonical ensemble. This is
\beq 
<E>=-\partial_{\be}\log Z=M
\label{mean}
\eeq
as it was expected, and the entropy is
\beq
S=\frac{A}{4}-\pi r_0^2(g-1)[1-\de(r_0,r_c)]
\label{entropy}
\eeq
where $A=r_+^2|\Im\tau|$ in the $g=1$ case. One can also derive these 
results from the on-shell Euclidean action, if only one takes the 
derivative of Eq.~(\ref{acT}) or Eq.~(\ref{acT1}) with respect to $T$.

\ss{Discussion}

From Eqs.~(\ref{mean}), (\ref{entropy}) it is clear how to proceed. 
First we notice that the entropy is exactly one quarter the area of 
the event horizon in the genus one black hole, the Hamiltonian mass is 
always positive and equal to the mean energy in the canonical 
ensemble. This is satisfactory and we shall not discuss this case any 
more. 

For the higher genus black holes, on the other hand, we have the 
choices $r_0=\ell\neq r_c$, which means the reference background is the 
$\eta=0$ solution of \AA minneborg et al., or $r_0=r_c=\ell/\sqrt{3}$, 
which is the zero temperature state corresponding to a naked 
singularity. We shall now discuss these two cases in order. 
The former choice was used in \cite{bril97-grqc,vanzo97-grqc} to define 
what is meant by the mass of the black holes. We stress that the 
prescription by which one defines the Hamiltonian mass and the 
Euclidean action should be the same, otherwise one runs into 
inconsistencies and, moreover, one cannot compare the two (see 
\cite{hawk96-13-1487} for a discussion of this relation between 
Hamiltonian mass and Euclidean action). Doing this consistently, we see from 
Eq.~(\ref{mass}) that there is a continuum of negative energy states in 
the range $-\ell/3\sqrt{3}\leq M<0$ (with positive specific heat, 
nevertheless), and the entropy picks up a topological contribution in the 
higher genus black holes (the sign of this was mistakenly taken to be 
positive in \cite{vanzo97-grqc}). The disaster is 
that the entropy becomes negative precisely when the temperature falls 
below the value $T_0=1/2\pi\ell$, at which the black hole's mass crosses zero 
becoming negative. A negative entropy does not make sense, so either 
one removes by hand the negative mass spectrum or 
interpret the negative entropy as meaning that negative energy states 
have an exponentially small probability of order $\exp(-3\pi(g-1)/|\La|)$. 
However, these putative negative mass black holes are perfectly 
acceptable solutions resembling much the Reissner-Nordstr\"{o}m 
solution, and have positive specific heat as the mass versus $T$ 
relation is concave upward everywhere. It seems rather arbitrary to 
cut them off, and we think, indeed, that this would be wrong. 

The second choice assign zero mass to the critical solution with 
$\eta_0=-\ell/3\sqrt{3}$, or $r_0=r_c$, which is not in fact a black 
hole (see \cite{bril97-grqc} for the causal Penrose diagram). Looking 
at the mass spectrum, Eq.~(\ref{mass}), we see it is positive for 
$T>0$ and zero only at $T=0$. Notice that the $\eta=0$ solution of 
\AA minneborg et al. has now positive mass, equal to $\ell(g-1)/\sqrt{27}$, 
but no curvature singularity at all. The near-to-zero temperature 
solution is a near-to-extreme black hole, which becomes a naked singularity 
at $T=0$ (the extreme Reissner-Nordstr\"{o}m solution, instead, is a 
black hole, but one on the verge of developing a naked singularity. It is 
only as a result of quantum emission that is driven away towards the 
non extreme solutions \cite{vanzo97-55-2192}). The third law of 
thermodynamics is thus perfectly consistent with the third law of 
black hole mechanics, according to which extreme solutions are 
forbidden. This is a satisfactory result, and we look now to the 
entropy of the black hole. 

Making the choice $\eta_0=-\ell/3\sqrt{3}$, which is equivalent to 
$r_0=r_c$, the unwanted negative term in the entropy formula 
disappears, leaving a positive definite entropy equal to one-quarter 
the area of the event horizon, in agreement with \cite{bril97-grqc} 
(which, however, have a 
partly negative mass spectrum since the $\eta=0$ solution was adjusted 
to zero mass). This indicates it is not $\eta$ that is related to the 
mass, but rather $\eta+\eta_0$ is. As discussed in the text, the reason 
for the disappearance of the unwanted term is quite subtle. The reference 
extreme background is the only solution, among the class considered, 
whose Euclidean section can be identified to any period in imaginary time 
without loosing the regularity of the metric. In all other backgrounds 
within the same topology class, there is a conic singularity even 
on-shell, which suddenly disappears in the extreme limit. 
We conclude that the topological black holes, at least in 
semiclassical quantum gravity, forms a well behaved sequence of 
positive mass solutions in anti-de Sitter gravity, with a stable 
thermodynamics. However, one cannot exclude the existence of 
topological transitions between different genus sectors. In this 
context, one may note that the genus of the black hole, or its conformal 
class in the toroidal case, is at this level a free, non dynamical parameter.  
This is unsatisfactory, and a better origin should be sought. 
As AdS is a possible ground state for string theory, it is 
not unlike that string theory and its parentage with the mathematics 
of the Riemann surfaces could do better than us. 

\ack{The author acknowledge useful conversations with S.~Zerbini, 
G.~Cognola and R.~Parentani.}


\begin{thebibliography}{10}

\bibitem{hawk73b}
{S.W.~Hawking and G.F.~Ellis}.
{\em {The large scale structure of spacetime}}.
{Cambridge University Press, Cambridge, England},  (1973).

\bibitem{hawk72-25-152}
{S. W.~Hawking}.
 Commun.~Math.~Phys. {\bf {25}}, {152} (1972).

\bibitem{frie93-71-1486}
{J.L.~Friedmann, K.~Schleich and D.M.~Witt}.
 Phys.~Rev.~Lett. {\bf {71}}, {1486} (1993).

\bibitem{jaco95-12-1055}
{T.~Jacobson and S.~Venkataramani}.
 Class. and Quantum Grav. {\bf {12}}, {1055} (1995).

\bibitem{shap95-52-6982}
{S.L.~Shapiro, S.A.~Teutolsky and J.~Winicour}.
Phys.~Rev. {\bf {D52}}, {6982} (1995).

\bibitem{gann76-7-219}
{D.~Gannon}.
 Gen.~Rel.~and Grav. {\bf {7}}, {219} (1976).

\bibitem{hawk77-15-2738}
{G.W.~Gibbons and S. W.~Hawking}.
 Phys.~Rev. {\bf {D15}}, {2738} (1977).

\bibitem{hawk83-87-577}
{S. W.~Hawking and D. N.~Page}.
 Commun.~Math.~Phys. {\bf {87}}, {577} (1983).

\bibitem{brow94-50-6394}
{J.D.~Brown, J.~Creighton and R. B.~Mann}.
 Phys.~Rev. {\bf {D50}}, {6394} (1994).

\bibitem{teit92-69-1849}
{M.~Ba$\tilde{n}$ados, C.~Teitelboim and J.~Zanelli}.
 Phys.~Rev.~Lett. {\bf {69}}, {1849} (1992).

\bibitem{bril96-69-249}
{D.~Brill}.
 Helv.~Phys.~Acta. {\bf {69}}, {249} (1996).

\bibitem{bana97-grqc}
{M.~Ba$\tilde{n}$ados}.
  {gr-qc/9703040} (1997).

\bibitem{lemo95-12-1081}
{J.P.S.~Lemos}.
 Class.~and Quantum Grav. {\bf {12}}, {1081} (1995).

\bibitem{lemo96-353-46}
{J.P.S.~Lemos}.
 Phys.~Lett. {\bf {B 353}}, {46} (1996).

\bibitem{stro91-360-197}
{G.T.~Horowitz and A.~Strominger}.
 Nuc.~Phys. {\bf {B360}}, {197} (1991).

\bibitem{lemo96-54-3840}
{J.P.S.~Lemos and V.T.~Zanchin}.
 Phys.~Rev. {\bf {D54}}, {3840} (1996).

\bibitem{peld96-13-2707}
{S.~\AA minneborg, I.~Bengtsson, S.~Holst and P.~Peldan}.
 Class.~Quantum Grav. {\bf {13}}, {2707} (1996).

\bibitem{mann97-14-L109}
{R.B.~Mann}.
 Class.~and Quantum Grav. {\bf {14}}, {L109} (1997).

\bibitem{vanzo97-grqc}
{L.~Vanzo}.
 {gr-qc/9705004}, (1997).

\bibitem{mann97-grqc}
{R.B.~Mann and W.L.~Smith}.
 {gr-qc/9703007} (1997).

\bibitem{fron65-37-221}
{C.~Fronsdal}.
 {Rev.~Mod.~Phys.} {\bf {37}}, {221} (1965).

\bibitem{avis78-18-3565}
{S.J.~Avis, C.J.~Isham and D.~Storey}.
 Phys.~Rev. {\bf {D18}}, {3565} (1978).

\bibitem{cole80-21-3305}
{S.~Coleman and F.~De Luccia}.
 {Phys.~Rev.} {\bf {D21}}, {3305} (1980).

\bibitem{brei82-144-249}
{P.~Breitenlohner and D.Z.~Freedman}.
 {Ann.~Phys.} {\bf {144}}, {249} (1982).

\bibitem{filt91-43-485}
{F.~Filthaut and C.~Dullemond}.
 {Phys.~Rev.} {\bf {D43}}, {485} (1991).

\bibitem{abbo82-195-76}
{L.F.~Abbot and S.~Deser}.
 {Nucl.~Phys.} {\bf {B195}}, {76} (1982).

\bibitem{frad87-177-63}
{E.S.~Fradkin and M.A.~Vasiliev}.
 {Ann.~Phys.} {\bf {177}}, {63} (1987).

\bibitem{frad91-261-26}
{E.F.~Fradkin and V.Ya.~Linetsky}.
 {Phys.~Lett.} {\bf {B261}}, {26} (1991).

\bibitem{louk96-54-2647}
{J.~Louko and S.N.~Winters-Hilt}.
 Phys.~Rev. {\bf {D54}}, {2647} (1996).

\bibitem{bril97-grqc}
{D.R.~Brill, J.~Louko and P.~Peldan}.
{gr-qc/9705012} (1997).

\bibitem{regg74-88-286}
{T.~Regge and C.~Teitelboim}.
 Ann.~of Phys. {\bf {88}}, {286} (1974).

\bibitem{york93-47-1407}
{J.D.~Brown and J.W.~York}.
 Phys.~Rev. {\bf {D47}}, {1407} (1993).

\bibitem{hawk96-13-1487}
{S.W.~Hawking and G.T.~Horowitz}.
 Class.~Quantum Grav. {\bf {13}}, {1487} (1996).

\bibitem{hayw93-47-3275}
{G.~Hayward}.
 Phys.~Rev. {\bf {D47}}, {3275} (1993).

\bibitem{mann97-05grqc}
{R.~Mann}.
{gr-qc/9705007} (1997).

\bibitem{hawk73-31-161}
{J.M.~Bardeen, B.~Carter and S.W.~Hawking}.
 Commun.~Math.~Phys. {\bf {31}}, {161} (1973).

\bibitem{gibb77-15-2752}
{G.W.~Gibbons and S.W.~Hawking}.
 Phys.~Rev. {\bf {D15}}, {2752} (1977).

\bibitem{teit94-51-4315}
{C.~Teitelboim}.
 Phys.~Rev. {\bf {D51}}, {4315} (1995).

\bibitem{suss94-50-2700}
{L.~Susskind and J.~Uglum}.
 Phys.~Rev. {\bf {D50}}, {2700} (1994).

\bibitem{furs96-54-2711}
{V.P.~Frolov, D.V.~Fursaev and A.I.~Zelnikov}.
 Phys.~Rev. {\bf {D54}}, {2711} (1996).

\bibitem{vanzo97-55-2192}
{L.~Vanzo}.
 Phys.~Rev. {\bf {D55}}, {2192} (1997).

\end{thebibliography}
\end{document}